\documentstyle[12pt,epsfig,psfig]{article}
\newcommand{\be}{\begin{equation}}
\newcommand{\ee}{\end{equation}}
\newcommand{\bea}{\begin{eqnarray}}
\newcommand{\eea}{\end{eqnarray}}

\begin{document}
\begin{flushright}
UFTP--492/1999\\
\end{flushright}
\vspace*{2cm}
\begin{center}
{\Large\bf 
Antiflow of Nucleons at the Softest Point of the EoS}
\\[2cm]
{\bf J.\ Brachmann, S.\ Soff, A.\ Dumitru,
H.\ St\"ocker, J.A.\ Maruhn, W.\ Greiner}
\\[0.4cm]
{\small Institut f\"ur Theoretische Physik der 
J.W.\ Goethe--Universit\"at}\\
{\small Postfach 111932, D--60054 Frankfurt a.M., Germany}
\\[0.7cm]
{\bf D.H.\ Rischke}
\\[0.4cm]
{\small RIKEN-BNL Research Center} \\
{\small Brookhaven National Laboratory, Upton, New York 11973, U.S.A.}
\\[1cm]
{\large July 1999}
\end{center}
\vspace*{1.5cm}
\begin{abstract}
We investigate flow in semi-peripheral nuclear collisions at AGS and SPS
energies
within macroscopic as well as microscopic transport models.
The hot and dense zone assumes the shape of an ellipsoid which
is tilted by an angle $\Theta$ with respect to the beam axis.
If matter is close to the softest point of the equation of state,
this ellipsoid expands predominantly {\em ortho\-gonal\/} to
the direction given by $\Theta$.
This antiflow component is responsible for 
the previously predicted reduction of the directed transverse momentum
around the softest point of the equation of state.
\end{abstract}

Transverse collective flow in relativistic nuclear collisions
reveals the properties of the nuclear matter equation of state
far from the ground state~\cite{oldflow,Yaris,Staubo:1990dq,SSoff2,Ollitraut,
Sorge,Danielnew,UH22,paper1}. 
In principle, one can distinguish three different types of transverse
collective flow: {\em radial}, {\em directed}, and {\em elliptic\/} 
\cite{Ollitraut}. Recent data on directed and elliptic flow
\cite{Harry,WA98,QM98,Pink} has revived theoretical interest
in this subject \cite{Sorge,Danielnew,UH22}.

Directed flow occurs only in semi-peripheral nuclear collisions, and
therefore must be studied in fully 3+1-dimensional geometries.
The beam axis is in general taken to be the $z$-direction, 
and the reaction plane to be the $z-x$--plane.
At BEVALAC energies, the two nuclei ``bounce off'' each other,
giving rise to a positive average momentum $\langle p_x(y)/N \rangle$ per 
nucleon in the forward direction \cite{oldflow}. In momentum space, 
the flow of matter can be described in terms of an ellipsoid,
defined by the principal axis' of the tensor of inertia~\cite{Daniel,oldflow},
which is tilted in the reaction plane by an angle $\Theta_{\rm flow}$
with respect to the beam axis. However, the actual shape of the distribution
of matter in momentum space needs not be ellipsoidal, see below.

In this paper we show that the situation is fundamentally different
if the equation of state of nuclear matter is softened, either by
a phase transition to the quark-gluon plasma or by the creation
of resonances and string-like excitations. To this end, we employ
one- \cite{Yaris,Staubo:1990dq} and three-fluid dynamics \cite{paper1},
as well as the microscopic model UrQMD \cite{UrQMD}.
We demonstrate that, around AGS energies,
the event shape resembles an ellipsoid in {\em coordinate\/} space,
tilted by an angle $\Theta$ with respect to the beam axis.
This ellipsoid expands predominantly {\em ortho\-gonal\/} to the
direction given by $\Theta$; we therefore term this flow component
{\em antiflow}. Around midrapidity, the antiflow largely cancels the
directed flow from the ``bounce--off'' of the two nuclei \cite{Csernai}.
We emphasize that here ``antiflow'' does not mean
the flow of antiparticles \cite{Jahns}, which is an absorption phenomenon,
nor the low energy ({\it i.e.} $E_{\rm Lab}^{\rm kin}
\simeq 100$~MeV/N) antiflow due to attractive potentials \cite{SSoff}.

This antiflow component has impact on studies of transverse elliptic flow
within simplified geometrical overlap models \cite{Ollitraut,UH22,VolPosnew}.
These studies assume that the longitudinal flow vanishes at $z=0$
in the {\em whole\/} transverse plane. 
The non-trivial ellipsoidal event shape, however, 
couples longitudinal to transverse flow, and
the longitudinal flow no longer vanishes everywhere
in the transverse plane at $z=0$. 
The amount of longitudinal flow 
is sensitive to the equation of state, as
well as the impact para\-meter and the bombarding energy, and
can only be determined in fully 3+1-dimensional calculations.

In order to measure the EoS, {\it i.e.}, in fluid-dynamical terms the pressure
$p(e,\rho)$ as a function of energy density $e$ and 
baryon density $\rho$ in the local rest frame of a fluid element, 
one studies the transverse momentum in the reaction plane, $p_x$.
This quantity is proportional to the pressure created in
the hot and dense collision zone \cite{oldflow}:
\be \label{eq:px}
p_x \sim \int p\, A_\perp \, {\rm d}t \quad .
\ee
The pressure $p$ is exerted over a transverse area $A_\perp$.
For increasing bombarding energy, the flow, $\sim p_x$, first
increases, as the compression and thus the pressure grow.
However, at large $E_{\rm Lab}^{\rm kin}$ the time span
of the collision decreases, diminishing the flow again. The
flow is thus maximized at some intermediate bombarding energy.

A phase transition softens the EoS \cite{Yaris}.
The pressure increases slower with $e$ and $\rho$ than in the case without
phase transition, reducing the velocity of sound.
This delays the fluid-dynamical expansion considerably, giving
the spectators time to pass the hot and dense zone, before they are
deflected.
One-fluid calculations \cite{Yaris} therefore show a local minimum 
(at $\simeq 8$~AGeV) of the excitation function of the
directed flow per nucleon, defined as
\be
\label{eq:flowdef}
\frac{p_x^{\rm dir}}{N} \equiv \left(\int{\rm d}y \, \frac{{\rm d}N}{{\rm d}y}
\right)^{-1}
\int{\rm d}y\, \frac{{\rm d}N}{{\rm d}y} \,  \left\langle 
\frac{p_x}{N} (y) \right\rangle  \, {\rm sgn}(y) \quad .
\ee
This is the weighted mean transverse in-plane
momentum $\langle p_x/N(y)\rangle$ per nucleon, introduced in \cite{Daniel}.
The weight is the net-baryon rapidity distribution, ${\rm d}N/{\rm d}y$.
In a fluid-dynamical context, 
the mean transverse momentum $\langle p_x/N(y)\rangle$ is defined as
\be \label{eq:pxN}
\left\langle \frac{p_x}{N}(y)\right\rangle = 
\frac{\int_y \!{\rm d}^3{\bf x} \,
R({\bf x})\,m_N\,u_x({\bf x})}
{\int_y \!{\rm d}^3{\bf x} \, R({\bf x})}\,\, ,
\ee
$u_x\equiv\gamma\, v_x$ denotes the $x$--component of the local 
4-velocity field, and $m_N$ is the nucleon rest mass.
$R$ is the zero-component of the net-baryon 4-current,
$R=\gamma\rho$.
Here, thermal smearing is neglected, and it is assumed that the 
$x$--component of the nucleon momentum can be approximated by
$m_N\,u_x$.
The volume integration is performed over all fluid elements
(projectile and target) around a given rapidity $y$.\footnote{
In the three-fluid model, since the 
third fluid is by construction baryon-free, the integration
covers only projectile and target fluids.}

\begin{figure}[htb]
\vspace*{-0cm}
\centerline{\hspace{0cm}{\hbox{\psfig{figure=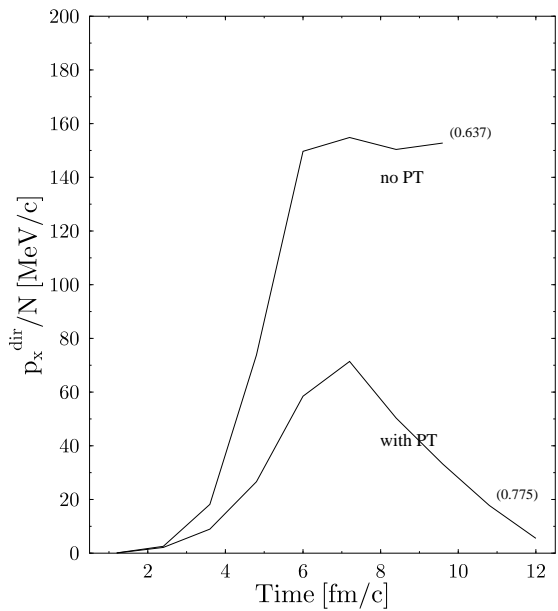,
height=8cm}}}}
\caption{Time-evolution (in the CM-frame)
of directed flow, $p_x^{\rm dir}/N$, for a $Au+Au$ reaction at
8~AGeV, $b=3$~fm, with and without 
phase transition to the QGP, calculated in one-fluid 
dynamics. The numbers in parentheses denote the mean
net-baryon density in units of the ground state density
$\rho_0 \simeq 0.16$ fm$^{-3}$ at the end of the 
time evolution.}
\label{fig:pxdir_time_1f}
\vspace*{.5cm}
\end{figure}
The EoS used in our one- and three-fluid calculations includes
a first order phase transition to a quark-gluon plasma (QGP). 
The hadronic phase consists of nucleons
interacting via exchange of $\sigma$ and $\omega$ mesons \cite{GorEoS},
and of non-interacting, massive pions. The QGP phase is described in
the framework of the MIT-Bag model \cite{MIT} 
as a non-interacting gas of massless $u$ and $d$ quarks 
and gluons, with a bag para\-meter $B^{1/4}=235$~MeV, resulting in a critical
temperature $T_c\simeq 170$~MeV. There is a first
order phase transition between these phases, constructed via
Gibbs' conditions of phase coexistence.

In Fig.\ \ref{fig:pxdir_time_1f}, we compute the
time evolution of the directed flow, $p_x^{\rm dir}/N$, 
in one-fluid dynamics, for a $Au+Au$ collision at impact para\-meter 
$b= 3$ fm and collision energy $E_{\rm Lab}^{\rm kin} = 8$~AGeV.
One observes that, due to the softening of the EoS in a phase
transition to the QGP, 
less directed flow is produced in the {\em early\/} compression
stage than in a purely hadronic scenario.
In contrast to the hadronic case, where the directed flow 
remains constant after reaching its maximum, in the case of
a phase transition, the directed flow decreases again.
By the time the mean density drops below nuclear ground-state density,  
$p_x^{\rm dir}/N$ is reduced to $\simeq 0$~MeV. If one follows
the fluid evolution even further (to unphysically small values of
the density), $p_x^{\rm dir}/N$ becomes negative.

\begin{figure}[htb]
\centerline{\hspace{0cm}{\hbox{\psfig{figure=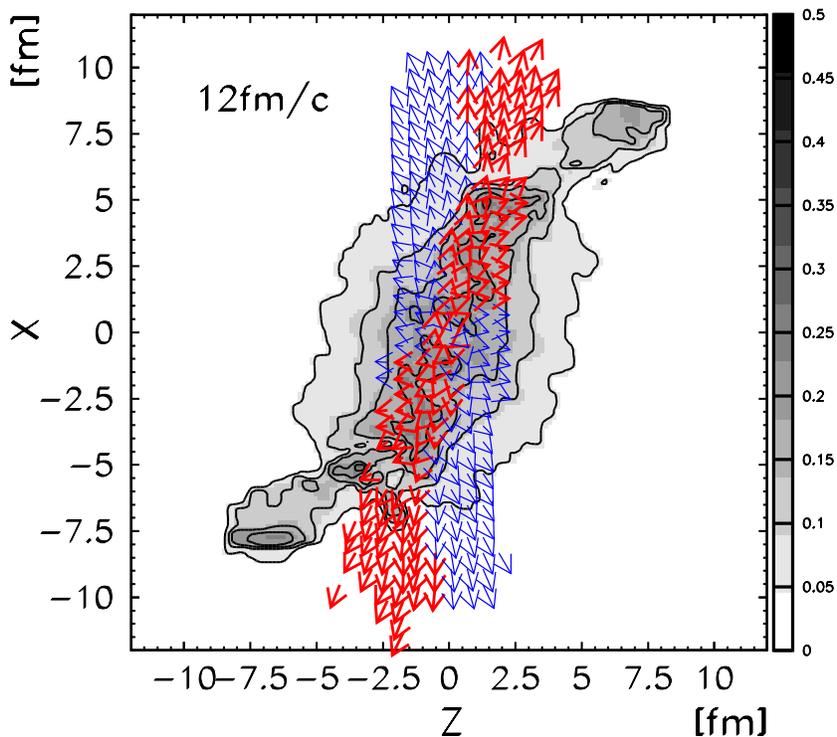,width=11cm}}
}}
\caption{Net-baryon density $R$ (for the same reaction as in
Fig.~\protect\ref{fig:pxdir_time_1f})
at $t=12$~fm/c in the reaction plane with
velocity arrows for midrapidity ($|y|<0.5$) fluid elements:
Antiflow - thin arrows, Normal flow - bold arrows.}
\label{fig:reacplaneA}
\vspace*{.5cm}
\end{figure}
This observation is explained by an
{\em antiflow\/} component which develops when the expansion sets in. 
This phenomenon is shown in Fig.\ \ref{fig:reacplaneA}, which is a contour
plot of the baryon density $R$, with arrows indicating the
fluid velocity. Normal flow (bold arrows) is {\em positive\/} in the
{\em forward\/} hemisphere, and negative in the backward hemisphere,
respectively. On the other hand, antiflow (thin arrows) is {\em positive\/}
in the {\em backward\/} hemisphere, and negative in the forward direction.
We show velocity arrows for fluid elements within $\pm 0.5$ units around 
midrapidity, since this phenomenon develops at midrapidity, as
discussed in detail below.
Similar results have been reported in \cite{LariAntiflow} within the 
microscopic quark-gluon string model \cite{QGSM}. 

In Fig.\ \ref{fig:reacplaneA} one observes that
the hot and dense zone assumes the shape of an ellipsoid tilted with
respect to the beam axis by an angle $\Theta$. The ellipsoid expands 
preferentially in the direction where its surface area is largest,
cf.\ (\ref{eq:px}), {\it i.e.}, {\em ortho\-gonal\/} to the direction
of the normal flow. This causes the antiflow.
Moreover, expansion into the direction of normal flow
is blocked by the spectators. (Similar arguments led to the prediction of
in-plane elliptic flow at high bombarding energies \cite{Ollitraut,
Sorge,Harry}.) Note that, at $z=0$, antiflow has a negative
longitudinal component for $x>0$, and a positive component for
$x<0$. This is the aforementioned coupling of longitudinal and
transverse flow in the central plane.

\begin{figure}[htp]
\centerline{\hspace{0cm}{\hbox{\psfig{figure=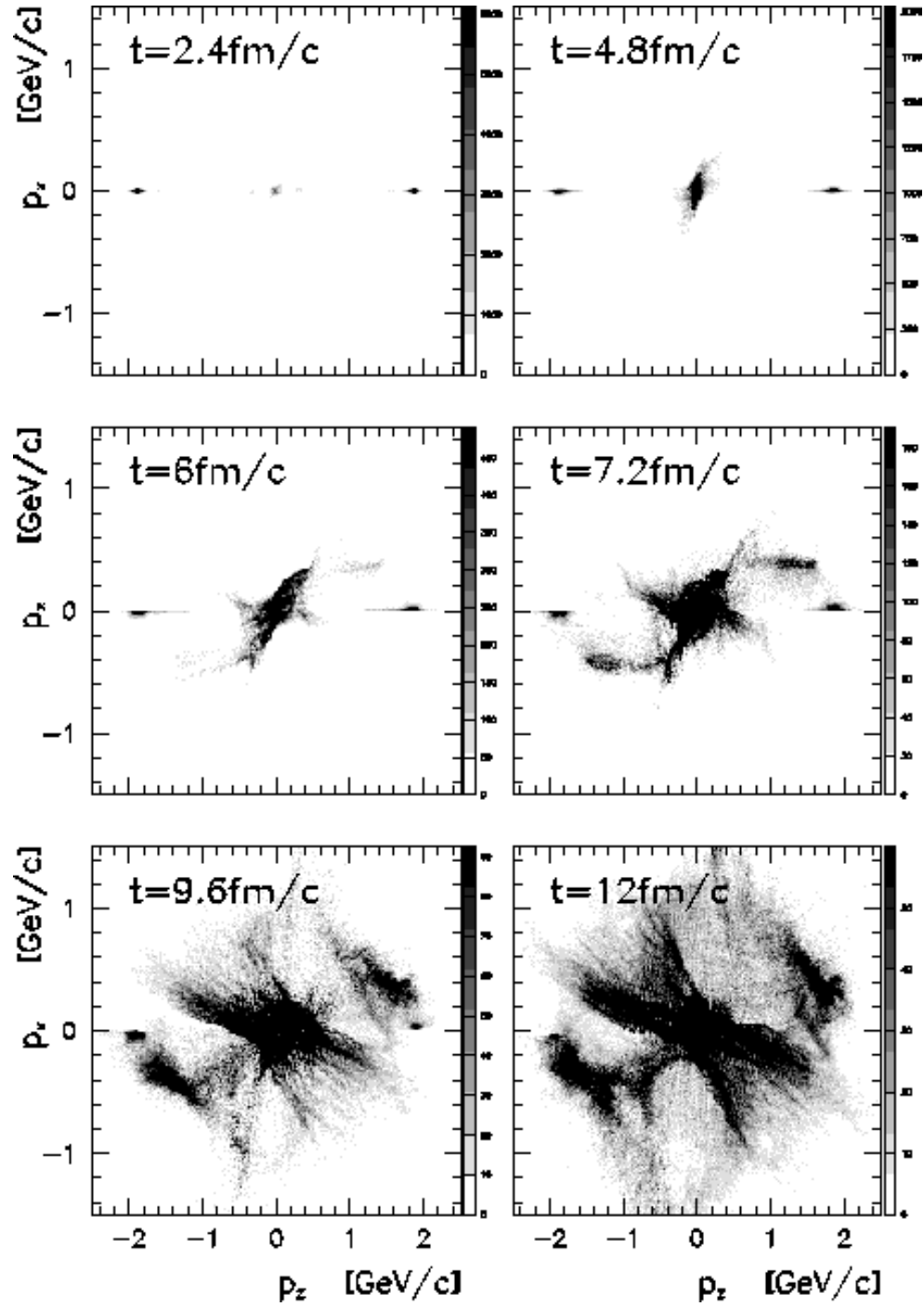,width=13cm}}
}}
\caption{Evolution of the net-baryon number in momentum space
within one-fluid dynamics (for the same reaction as in
Fig.~\protect\ref{fig:pxdir_time_1f}).}
\label{fig:pxpz}
\vspace*{.5cm}
\end{figure}
The evolution of the distribution of nucleons in momentum space is depicted
in Fig.~\ref{fig:pxpz}. In this one-fluid calculation, the participants are
shifted instantaneously to midrapidity. 
In the early stage, $t\simeq2.4$~fm/c, they can be found
around ${\bf p}\sim0$. (For clarity,  the Fermi-momentum and the
thermal momenta of the nucleons in the local rest-frame of the fluid are
not included.) At $t\simeq4.8$~fm/c the normal flow builds up around
central rapidities, leading to an 'ellipsoidal' distribution. The
principal axis is tilted with respect to the rapidity axis by
$\Theta_{\rm flow}\simeq\pi/4$. However,
at even later times, an additional orthogonal component, the anti-flow,
builds up. This is due to the expansion of the above-mentioned ellipsoid
in {\em coordinate} space, which proceeds in the direction of maximal
surface. The final distribution in momentum space can even be dominated by
the anti-flow component and therefore does not exhibit an ellipsoidal
shape.

\begin{figure}[hpt]
\centerline{\hspace{0cm}{\hbox{\psfig{figure=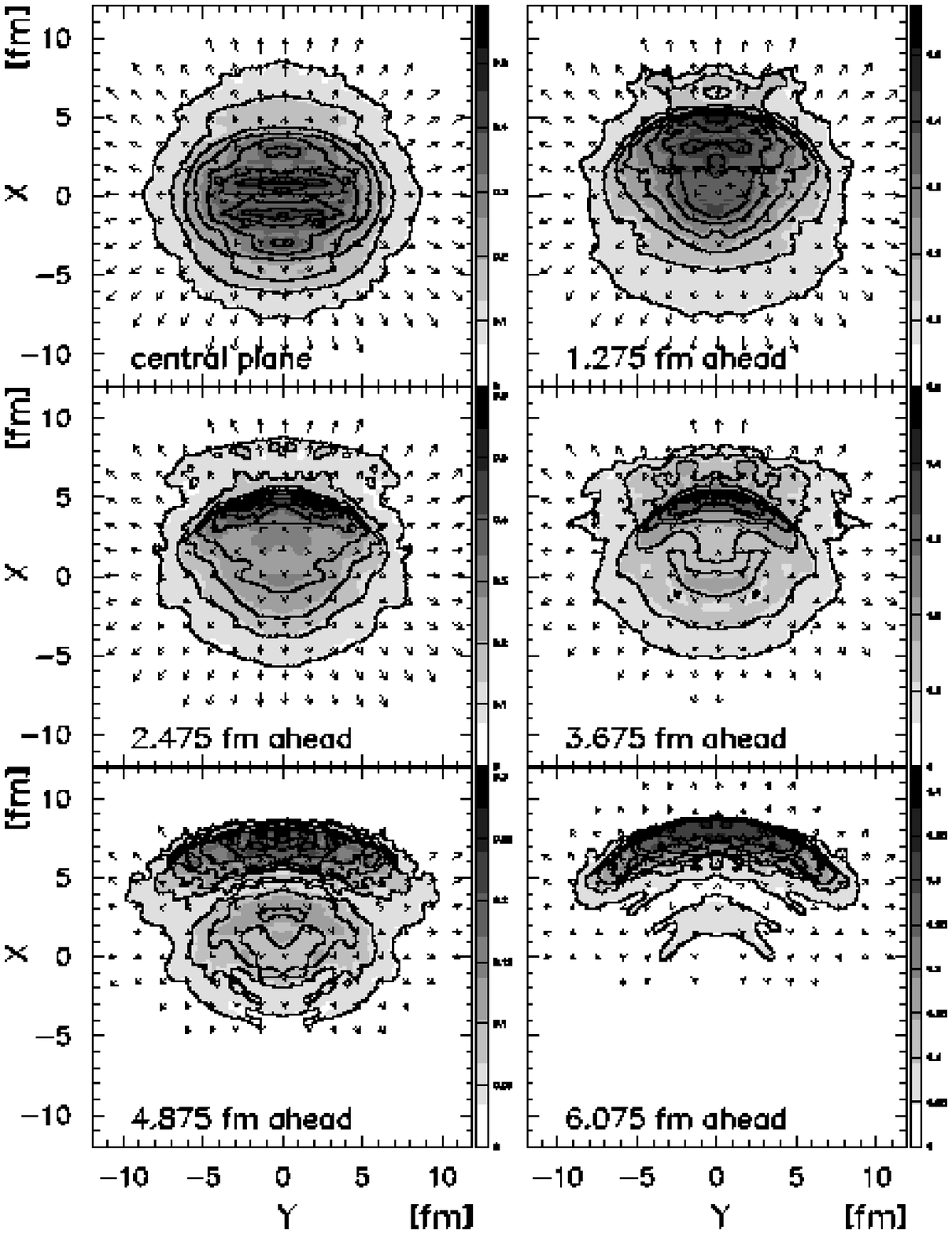,width=13cm}}}}
\caption{Contour plots of $R$ in the tranverse $x-y$--plane at
several values of $z>0$ (forward hemisphere).}
\label{fig:perplane}
\vspace*{.5cm}
\end{figure}
To illustrate the three-dimensional structure of the expanding matter in
coordinate space, we also show the baryon density distribution
in the transverse plane
at various values of $z$. The system in the central plane, at $z=0$
(upper left panel) is symmetric around the reaction plane, confirming
the assumption made in simple geometrical overlap models used
to study elliptic flow (however, as discussed above, the
longitudinal flow does {\em not\/} vanish at $z=0$).
Furthermore, for $z>0$ (forward, or projectile, hemisphere)
the system is displaced towards positive $x$. 
For $z=1.275$ and $2.475$~fm, antiflow is clearly visible as
flow of matter towards negative $x$.
In the most forward plane, $z= 6.075$ fm, only spectators remain,
which are ''cut off'' from the central region and flow mainly in the positive
$x$--direction.

We have also studied directed flow in the
three-fluid model, with a dynamical local unification procedure.
The three-fluid model \cite{paper1} treats the nucleons of the projectile and
target nuclei as two different fluids, since they populate different
rapidity regions in the beginning of the reaction. The same holds for the
newly produced particles around midrapidity, which are therefore collected
in the third fluid. Thus, the three-fluid model accounts for the 
non-equilibrium situation during the compression stage of 
heavy-ion collisions. The coupling between the projectile and target fluids 
leads to a gradual deceleration and is parametrized by 
free binary $NN$-collisions \cite{2fluid}.

The unification of fluids $i$ and $j$ consists of adding their
energy-momentum tensors and net-baryon currents in the respective
cells,
\be
T_i^{\mu \nu}(x) + T_j^{\mu \nu}(x) = T^{\mu \nu}_{\rm unified}(x)
\;\;\;\; ,\;\;\;\;\; N^\mu_i(x) + N^\mu_j(x) = N^\mu_{\rm unified}(x)
\ee
and common values for $e,\, p,\,\rho$ and $u^\mu$ are obtained
from $T^{\mu \nu}_{\rm unified} = (e+p)\, u^\mu u^\nu - p \, g^{\mu\nu}$,
$N^\mu_{\rm unified}  = \rho\, u^\mu$, and the given EoS $p=p(e,\rho)$.
The local criterion for
unification is
\be
\label{eq:onef_crit}
\frac{p_i+p_j}{p} > 0.9 \quad .
\ee
Here, $p_{i,j}$ denotes the pressure in $T^{\mu \nu}_{i,j}$,
and $p$ the pressure in $T^{\mu \nu}_{\rm unified}$.
Eq.~(\ref{eq:onef_crit}) has already been used in \cite{paper1} 
as a measure for the equilibration process.

\begin{figure}[htb]
\centerline{\hspace{0cm}{\hbox{
\psfig{figure=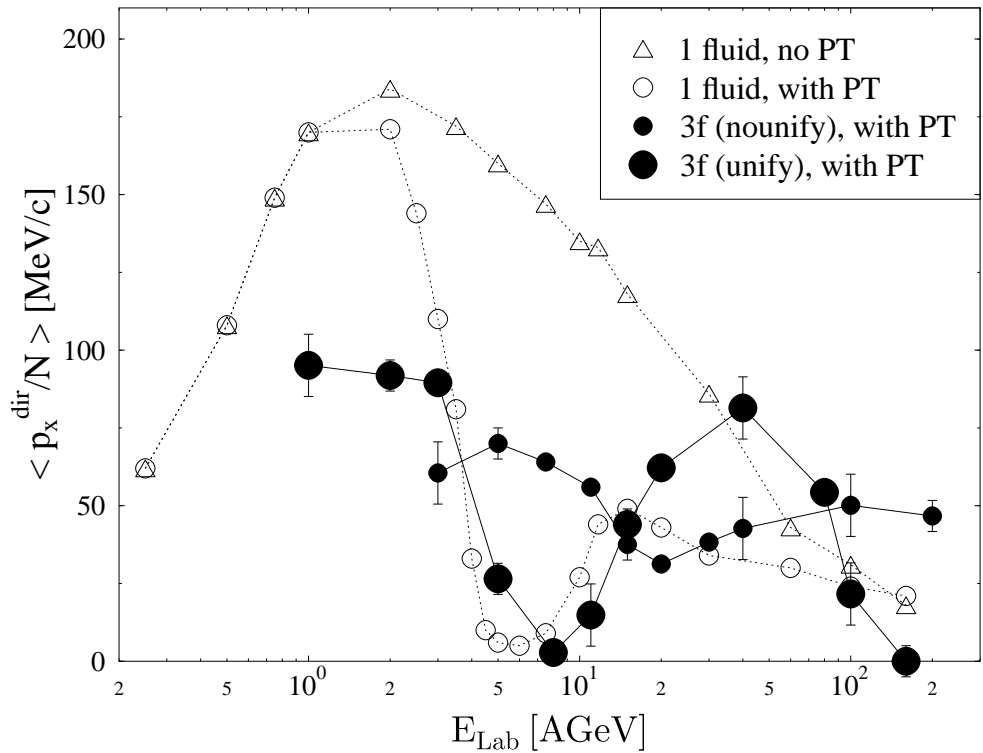,angle=-90,width=13cm}}}}
\caption{The excitation function of directed flow $p_x^{\rm dir}/N$ for 
$Au+Au$ collisions at impact para\-meter $b=3$~fm. Dotted lines are 
results from a one-fluid calculation; triangles are for a purely
hadronic EoS, circles are for an EoS with phase transition. 
Solid lines are calculated with the three-fluid model, with (large circles)
or without (small circles) dynamical unification.
All three-fluid calculations are performed with an EoS 
with phase transition.}
\label{fig:pxy_3funify}
\vspace*{.5cm}
\end{figure}

Fig.\ \ref{fig:pxy_3funify} shows the excitation function of directed flow
$p_x^{\rm dir}/N$ calculated in the three-fluid model in comparison
to that obtained in a one-fluid calculation \cite{Yaris}.
Due to non-equilibrium effects in the early stage of the reaction,
which delay the build-up of transverse pressure \cite{Sorge,paper1},
the flow in the three-fluid model is reduced as compared to the
one-fluid calculation in the AGS energy range. Furthermore, the 
minimum in the excitation function of the directed flow
shifts to higher bombarding energies.
The case without dynamical unification yields
the least amount of stopping and energy deposition, while
the one-fluid calculation has instantaneous
full stopping and maximum energy deposition.
The three-fluid model with dynamical unification lies between these
two limits; it accounts for the limited stopping power of nuclear matter
in the early stages of the collision and mutual equilibration
of the different fluids in the later stages.
Consequently, the shift of the minimum is large without,
and rather moderate with unification.

The three-fluid model predicts a local minimum in
the excitation function of directed flow at bombarding energies
between $10$ and $20$~AGeV, depending on the fluid unification criterion
(\ref{eq:onef_crit}). While
measurements of flow at AGS energies \cite{QM98} have
found a decrease of directed flow with increasing
bombarding energy, a minimum has so far not been observed.
In the three-fluid model with unification, the directed flow
exhibits a local maximum at $E_{\rm Lab}^{\rm kin} \sim 40$~AGeV.
If recent CERN-SPS experiments \cite{stock} find larger values for
the directed flow than at the maximum AGS energy, the existence
of a minimum in the excitation function of the directed flow
due to the intermediate softening of the EoS would be unambigously proven.

We emphasize that the excitation function depicted in Fig.~\ref{fig:pxy_3funify}
has been calculated for fixed impact para\-meter $b$, which
is not directly measurable in an experiment.
Usually the amount of transverse energy or the number of participating nucleons
are employed as measures for $b$, assuming that
the interaction volume is given by the geometrical
overlap of two spheres displaced in $x$-direction by the amount $b$.
However, the above discussion suggests that such a geometry is
oversimplified. The two nuclei are partly deflected and stopped, and thus
do not penetrate as deeply as compared to the simple geometrical overlap case.
Furthermore, it is also not obvious that the same $E_t/E_t^{max}$-bin at different
bombarding energies corresponds to the same impact para\-meter, since the system geometry
may change considerably due to energy-dependent phenomena like the stopping power,
phase transitions, etc.
This should be kept in mind when considering the different values of $b$
where the directed flow is strongest: ${b_m\approx 4}$~fm at AGS \cite{AGSdata} and
${b_m\approx 8}$~fm at SPS \cite{WA98}.
A detailed study of the impact para\-meter dependence of directed flow,
transverse energy production and number of participating nucleons
within the three-fluid model is in preparation.

\begin{figure}[htp]
\centerline{\hspace{0cm}{\hbox{\psfig{figure=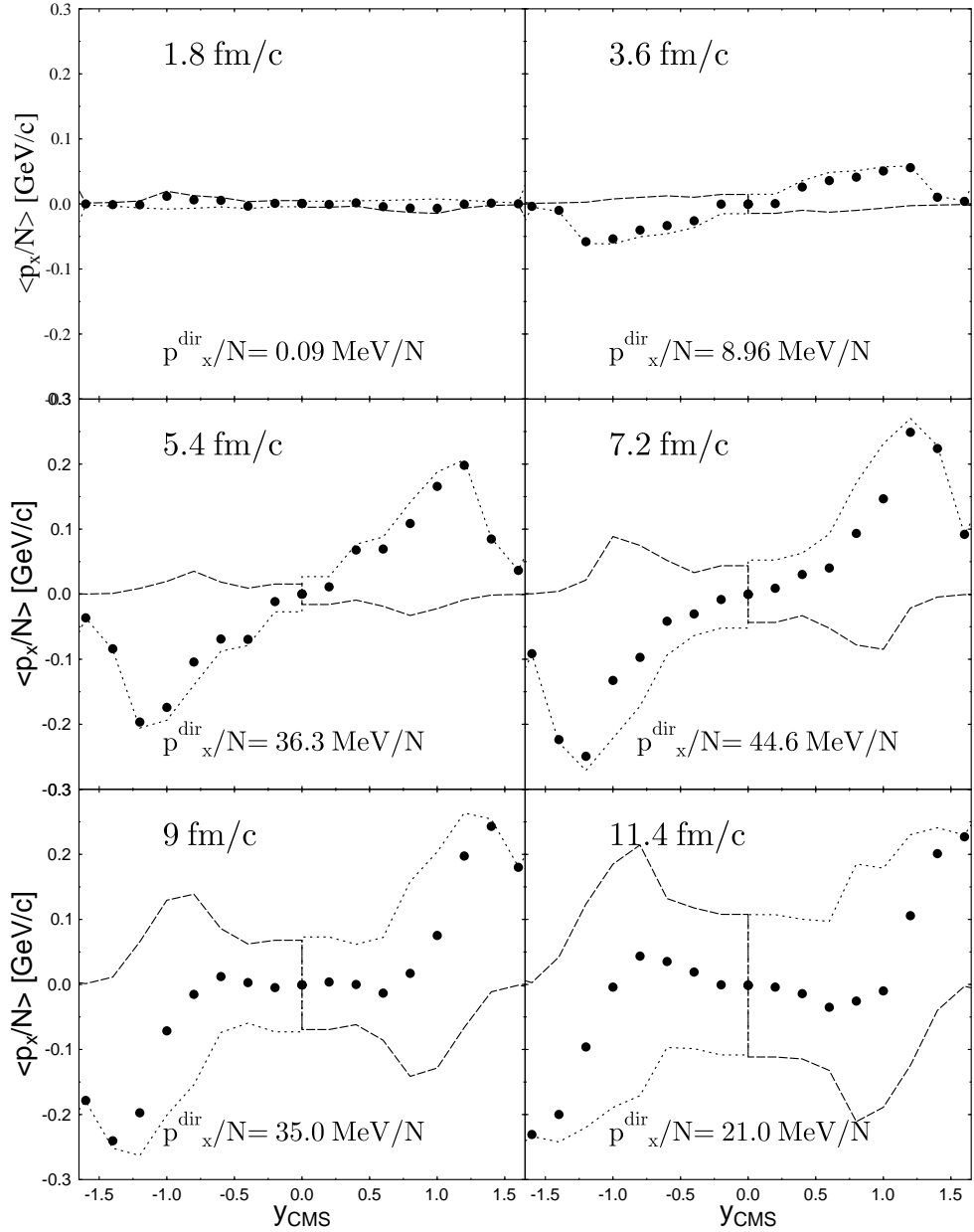,width=13cm}
}}}
\caption{Mean in-plane momentum per nucleon $\langle p_x/N(y) \rangle$ 
as function of rapidity $y$ at various times (dots). 
The dotted and dashed lines 
show the decomposition into flow and antiflow, respectively. }
\label{fig:pxy_11gev}
\vspace*{.5cm}
\end{figure}

Let us return to the discussion of the antiflow, which develops also in the
three-fluid model at energies around the minimum in the excitation function of 
$p_x^{\rm dir}/N$. It leads to a plateau
in $\langle p_x/N(y) \rangle$ around midrapidity. 
In Fig.\ \ref{fig:pxy_11gev}, this quantity is shown as a
function of rapidity $y$ at different times. Observe that in the late stage
of the reaction, close to freeze-out, the flow around $y=0$ is even
{\em negative}.

We locally decompose the flow into a normal component and 
an antiflow component
\bea \label{flowdecomp}
    \mbox{normal flow} &:& y({\bf x}) \, p_x({\bf x}) > 0 \,\, ,\\
    \mbox{antiflow} &:& y({\bf x}) \, p_x({\bf x}) < 0 \,\, .
\eea
Consequently, we define
\be \label{eq:pxNflowAflow}
\left\langle \frac{p_x^{\rm flow/antiflow}}{N}(y) \right\rangle
\equiv \frac{\int_y \!{\rm d}^3{\bf x} \,
R({\bf x})\,m_N\,u_x({\bf x}) \, \theta \left[ \pm y({\bf x}) \,
p_x({\bf x})\right] }{\int_y \!{\rm d}^3{\bf x} \, R({\bf x})} \,\, .
\ee
The individual components $\langle p_x^{\rm flow}/N(y)\rangle$ 
and $\langle p_x^{\rm antiflow}/N(y)\rangle$ are 
also shown in Fig.\ \ref{fig:pxy_11gev}.
The antiflow component develops from midrapidity after $\simeq 6$~fm/c.
According to the definition~(\ref{eq:pxNflowAflow}), both the normal
flow and the antiflow are discontinuous at $y=0$.
The sum of both components yields the total flow, eq.\ (\ref{eq:pxN}),
which is continuous (and equal to zero) at midrapidity.

\begin{figure}[htb]
\centerline{\hspace{0cm}{\hbox{\psfig{figure=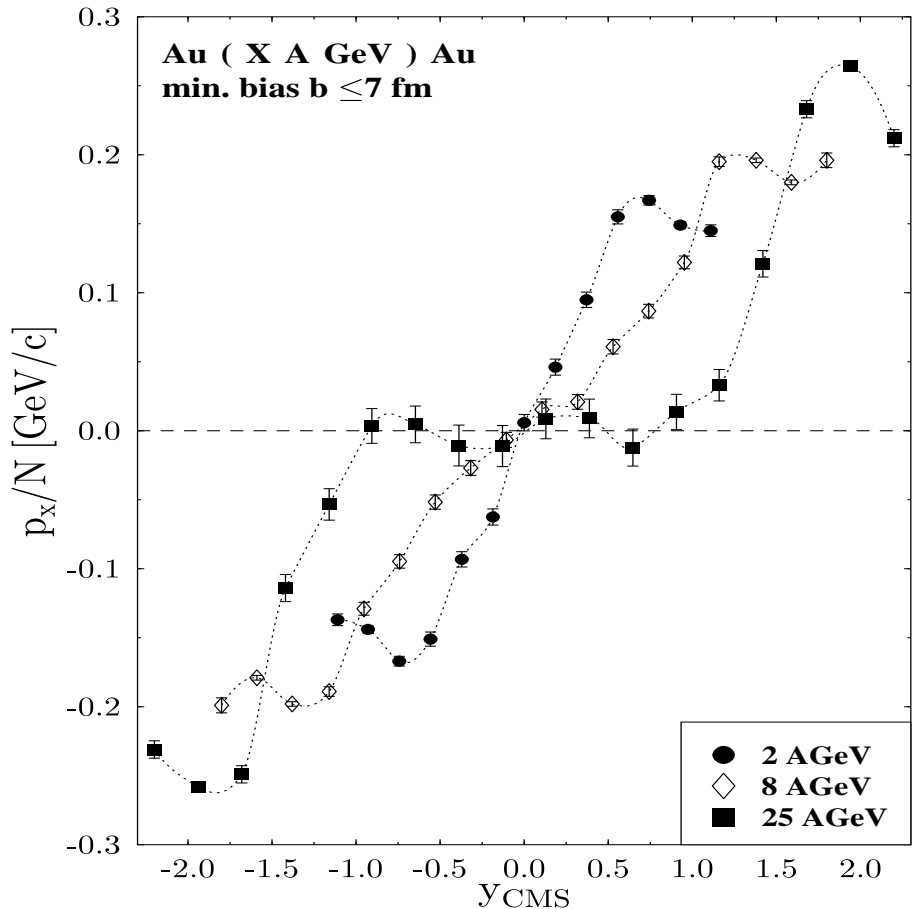,width=12cm,
height=12cm}}}}
\caption{In-plane transverse momentum distributions at various
energies as calculated in the UrQMD model. Calculations were performed using
density-dependent Skyrme potentials corresponding to a hard EoS
with an incompressibility ${K \approx 380}$~MeV.}
\label{fig:pxy_urqmd}
\vspace*{.5cm}
\end{figure}

This phenomenon is not only limited to fluid-dynamical models.
The plateau around midrapidity in $\langle p_x/N(y) \rangle$ 
is also visible in the microscopic UrQMD model \cite{UrQMD}. 
Fig.\ \ref{fig:pxy_urqmd} shows the respective
$\langle p_x/N(y) \rangle$ for $Au+Au$ collisions 
at various bombarding energies.
The flattening around midrapidity is more pronounced
at larger energy.
A similar behavior has been found in other microscopic models
\cite{LariAntiflow,Liu}. Unlike the fluid-dynamical calculations,
the microscopic transport model does not really show a negative slope of
$\langle p_x/N(y) \rangle$ around midrapidity.
Unfortunately, measurements of $\langle p_x/N(y) \rangle$ 
at beam energies $E=6-11$~AGeV cover only the range
$y \geq 0.3\, y_{P,CMS}$ \cite{AGSdata2,Liu}. The quantitative
value of antiflow at midrapidity thus remains undetermined.

\begin{figure}[htp]
\centerline{\hspace{0cm}{\hbox{\psfig{figure=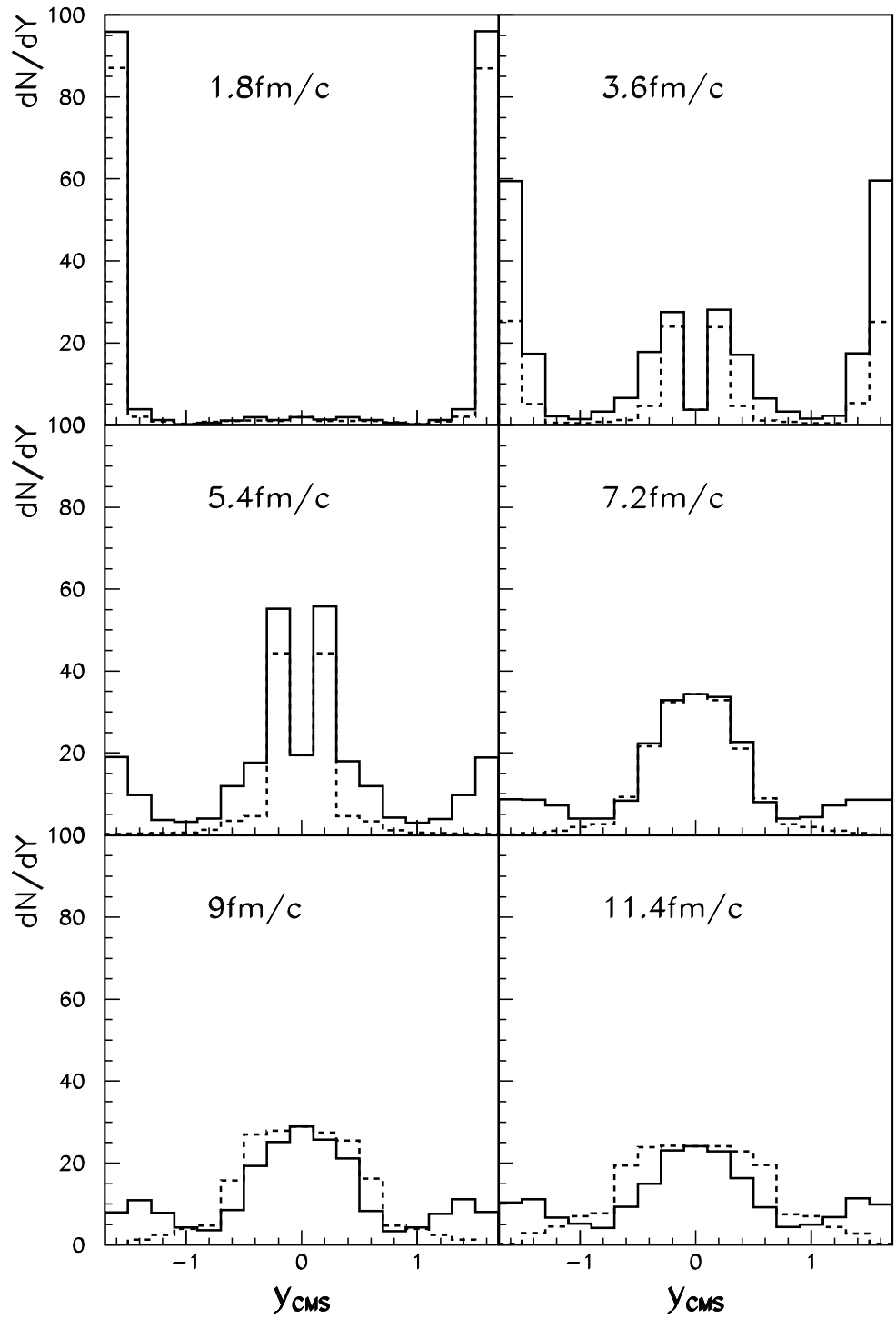,width=13cm}
}}}
\caption{Rapidity distributions of normal flow (full line) and antiflow
(dashed line) at various times.}
\label{fig:dndy_posnegall}
\vspace*{.5cm}
\end{figure}
Besides the transverse directed momentum $p_x^{dir}/N$, many other 
measures for the in-plane flow exist, for instance, 
the slope of $ \langle p_x/N(y) \rangle$ or 
$\langle p_x/N(y/y_{\rm beam})\rangle$ at midrapidity, or their maximum
values as function of $y$, or their values at $y_{\rm beam}/2$.
All these measures will yield qualitatively and quantitatively different 
results, because flow and antiflow develop in different rapidity regions,
as seen in Fig.\ \ref{fig:dndy_posnegall}.
In the early compression stage of the reaction the spectators near 
projectile and target rapidity are deflected by the pressure in the 
central hot and dense zone, producing normal flow
(Fig.\ \ref{fig:dndy_posnegall}, full line).
When the expansion of the hot and dense zone sets in, 
both normal flow and antiflow (Fig.\ \ref{fig:dndy_posnegall}, dashed line) 
develop around midrapidity. 
Nevertheless, the antiflow finally occupies a broader region
around midrapidity than the normal flow, while the normal flow dominates
the region near projectile and target rapidities. 

In summary, we investigated transverse directed flow in macroscopic as well as
microscopic transport models. For the three-fluid model,
we find a minimum in the excitation function of the directed flow
at bombarding energies between $10$ and $20$~AGeV, which is
somewhat above the value found earlier in one-fluid calculations.
The minimum is caused by a softening of the EoS due to a phase
transition to the QGP. An antiflow component was identified
as source for the reduction of the directed flow and discussed in detail.
We also found that the directed flow of nucleons increases again at higher bombarding
energy, leading to a maximum in the excitation function at energies
around $40$~AGeV. If the directed flow at this energy, currently
investigated by CERN-SPS experiments, proves
to be larger than at maximum AGS energies, the existence
of a minimum in the excitation function of the directed flow
would be unambigously proven.

\vspace*{.5cm}
\noindent
{\large \bf Acknowledgements:}\\
This work was supported by DFG, BMBF, GSI and the Graduiertenkolleg
{\em Theoretische und Experimentelle Schwerionenphysik}.
We thank
L.\ Bravina for suggesting the decomposition~(\ref{flowdecomp}), and
L.\ Satarov and I.N.\ Mishustin for interesting discussions.
A.\ Dumitru also thanks T.\ Awes from the WA98 collaboration for helpful
discussions on their data.
We are grateful to M.\ Bleicher and S.A.\ Bass for reading the manuscript
prior to publication.

\end{document}